\documentclass[amsmath,
twocolumn,
aip,apl]{revtex4}
\usepackage{graphicx}% Include figure files
\usepackage{dcolumn}% Align table columns on decimal point
\usepackage[colorlinks=true, pdfstartview=FitV,
linkcolor=blue, citecolor=blue, urlcolor=blue]{hyperref}

\begin{document}

\title{A geometry for optimizing nanoscale magnetic resonance force
  microscopy}

\author{Fei Xue, P. Peddibhotla, M. Montinaro, D. Weber, and M. Poggio}

\affiliation{Department of
  Physics, University of Basel, Klingelbergstrasse 82, 4056 Basel,
  Switzerland}

\date{\today}

\begin{abstract}
 We implement magnetic resonance force microscopy (MRFM) in an
  experimental geometry, where the long axis of the cantilever is
  normal to both the external magnetic field and the RF microwire
  source. Measurements are made of the statistical polarization of
  $^1$H in polystyrene with negligible magnetic dissipation, gradients
  greater than $10^5$ T/m within 100 nm of the magnetic tip, and
  rotating RF magnetic fields over 12 mT at 115 MHz.  This geometry
  could facilitate the application of nanometer-scale MRFM to nuclear
  species with low gyro-magnetic ratios and samples with broadened
  resonances, such as In spins in quantum dots.

\end{abstract}


\maketitle

In the past several years, researchers have employed force-detected
nuclear magnetic resonance to extend the resolution of magnetic
resonance imaging (MRI) into the nanometer-scale \cite{Poggio:2010}.
In 2009, using a technique known as magnetic resonance force
microscopy (MRFM), researchers made 3D images of individual tobacco
mosaic viruses with a resolution better than 10 nm \cite{Degen:2009}.
The sensitivity of MRFM now surpasses the sensitivity of conventional,
inductively-detected magnetic resonance by 8 orders of magnitude,
allowing for the detection of small spin ensembles containing less
than $10^4$ nuclear moments.  This increased sensitivity has opened
the door for MRI of structures that, until now, had been inaccessible
to conventional techniques, e.g. individual virus particles, thin
films \cite{Mamin:2009}, and -- potentially -- quantum dots (QDs).

The most recent implementations achieving the highest sensitivity
\cite{Degen:2009,Mamin:2009,PoggioAPL:2007,Degen:2007,HighSensMRFM}
have utilized a sample-on-cantilever configuration with the sample
affixed to the end of an ultra-sensitive Si cantilever having a spring
constant typically on the order of 100 $\mu$N/m.  The sample is
positioned within 100 nm of a FeCo magnetic tip in a ``pendulum''
geometry so as to avoid snap in to contact \cite{Stowe:1997}.  The
magnetic tip is patterned on top of an RF microwire which acts as an
RF magnetic field source.

\begin{figure}[bp]
\includegraphics[trim=8.1cm 14.2cm 7.5cm 13.7cm,clip, width=8.5cm, page=1]{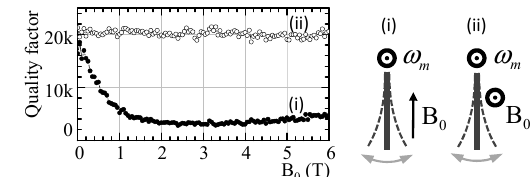}
\caption{
\label{fig1}
A nominally non-magnetic 120 $\mu$m $\times$ 4 $\mu$m $\times$ 0.1
$\mu$m Si cantilever \cite{Chui:2003}.  Quality factor $Q$ plotted as a function of
magnetic field pointing perpendicular (i) and parallel (ii) to the
lever's angular rotation vector at $T = 4.2$ K}
\end{figure}

One important limitation to the spin sensitivity is mechanical
dissipation experienced by the cantilever in a static magnetic field.
As MRFM experiments are carried out at elevated magnetic fields in
order to saturate the magnetic tips, magnetic dissipation in the
cantilever oscillator can significantly reduce the spin sensitivity of
a measurement. For mechanical oscillators containing magnetic
materials, dissipation that depends on magnetic field is expected and
has been observed \cite{Stipe:2001,Harris:2003}.  Magnetic dissipation
should not affect oscillators containing no magnetic material.
Recently, however, it has been reported that the application of a
magnetic field can cause significant mechanical dissipation in
micro-mechanical cantilevers made of nominally non-magnetic Si
\cite{Harrington:2000,Stipe:2001,Mohanty:2002,Ng:2006}.  As shown in
Fig.~\ref{fig1}(b), this kind of dissipation has also been observed as
a loss in mechanical quality factor $Q$ in some bare ultra-sensitive
Si cantilevers used for recent MRFM.  Its origin is likely the
unintentional presence of impurities or defects with magnetic
anisotropy, probably introduced during processing.

\begin{figure}[bp]
\includegraphics[trim=7.9cm 13.0cm 8.2cm 13.1cm,clip, width=8.5cm, page=2]{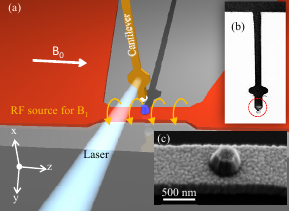}
\caption{ \label{fig2} (a) Experimental apparatus. Current flows in
  the microwire (red) along $\hat{z}$, while the lever displacement is
  along $\hat{y}$. Inset (b) shows an SEM image of an Si cantilever
  with a polystyrene sample and (c) shows a SEM of a microwire RF
  source with an integrated FeCo tip.}
\end{figure}

One way to eliminate this type of dissipation without having to
eliminate the presence of magnetic impurities is to ensure that as the
cantilever oscillates, it does not change its angle with respect to
the applied magnetic field ${\bf B}_0=B_0 \hat{z}$, as seen in
Fig.~\ref{fig1}.  We therefore show MRFM measurements of the
statistical polarization of $^1$H in a polystyrene particle where we
apply ${\bf B}_0$ along the cantilever's angular rotation vector.  As
shown in Fig.~\ref{fig2}, the force detection apparatus is similar to
the conventional apparatus described in \cite{PoggioAPL:2007} with the
important exception of the orientation of ${\bf B}_0$. A similar
geometry for MRFM was first proposed by Marohn et al.\ in 1998
\cite{Marohn:1998} in order to avoid magnetic spring effects and
magnetic dissipation.  Our apparatus, however, has the additional
advantage of being compatible with the microwire RF sources and
patterned magnetic tips designed for nanoscale MRFM.

The cantilever used in this experiment is 150 $\mu$m long, 4 $\mu$m
wide, and 0.1 $\mu$m thick and includes a 1-$\mu$m thick mass on its
end \cite{Chui:2003}. A droplet of polystyrene solution is affixed
and cured onto the end of the cantilever resulting in a 2-$\mu$m
sized particle \cite{SuppMat}. The sample-loaded cantilever has a
mechanical resonance frequency $\omega _{m} = 2 \pi \times 2.6$ kHz
and an intrinsic quality factor $Q = 5.0 \times 10^4$ at $T = 500$
mK. Through measurements of the cantilever's thermal motion, we
determine its effective spring constant to be $k = 120$ $\mu$N/m.
The microwire RF source and nano-sized magnetic tip are fabricated
in a process similar to that described in reference
\cite{PoggioAPL:2007}.  The Au wire is 2.6-$\mu$m long, 1.0-$\mu$m
wide, and 0.2-$\mu$m thick and is patterned atop a Si substrate. The
FeCo tip deposited atop the microwire is 250 nm tall in the shape of
a truncated cone with a diameter at the top and bottom being 270 nm
and 510 nm respectively. The MRFM apparatus is isolated from
vibrations and mounted in a vacuum chamber with a pressure below
$10^{-6}$ mbar at the bottom of a continuous flow $^3$He cryostat.
The motion of the lever is detected using 100 nW of 1550 nm laser
light focused onto a 10-$\mu$m wide paddle and reflected back into
an optical fiber interferometer \cite{Rugar:1989}.  We damp the
cantilever to a quality factor of $Q \approx 400$ in order to
increase the bandwidth of our force detection without sacrificing
force sensitivity \cite{Garbini:1996}.  The damping is realized
using an optimal-control feedback algorithm implemented on a
National Instruments FPGA board \cite{Jacky:2008}.

\begin{figure}[tp]
\includegraphics[trim=7.9cm 12.3cm 7.6cm 13.4cm,clip, width=8.5cm, page=3]{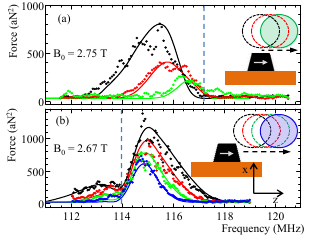}
\caption{ \label{fig3} $^1$H spin signal in a polystyrene sample as
a function of RF carrier frequency. Points are experimental data
while solid lines are simulations. The insets depict the positions
of the sample (colored circles) for each resonance curve relative to
the FeCo tip (magnetized along the white arrow). The dashed lines
mark the resonance frequencies of $^1$H spins at $B_0$.  }
\end{figure}

We measure the statistical polarization of $^1$H at $T = 500$ mK in
the polystyrene particle using the adiabatic rapid passage technique
described in reference \cite{PoggioAPL:2007, SuppMat}. Experiments
are performed with $B_0 > 2.5$ T such that the FeCo tip is fully
magnetized along $\hat{z}$.  At these fields, the cantilever
experiences no magnetic dissipation and maintains its intrinsic $Q$
of $5.0 \times 10^4$ far from the microwire surface.  During the
measurement, the sample at the end of the cantilever is positioned
within 200 nm of the FeCo tip.  Upon such close approach to the FeCo
tip and microwire, interactions between the sample and the microwire
surface begin to dominate the mechanical dissipation \cite{Surface};
these effects reduce the cantilever $Q$ to $1.5 \times 10^4$.  While
this kind of dissipation can be minimized by using the appropriate
coating on the end of the cantilever tip, it remains an important
limit on MRFM sensitivity.

Magnetic resonance measurements performed with the sample positioned
150 nm above the FeCo tip are shown in Fig.~\ref{fig3}(a).  The
lineshape of the magnetic resonance signal as a function of RF carrier
frequency is seen to change as the sample is moved from being directly
above the FeCo tip along $\hat{z}$ in steps of 240 nm.  A similar
experiment, in which the sample is positioned to one side of the FeCo
tip, such that its surface is about 80 nm from the closest point on
the FeCo tip, is shown in Fig.~\ref{fig3}(b).  Here the magnetic
resonance lineshape changes as the sample is moved away from the FeCo
tip along $\hat{z}$ in steps of 60 nm.  The data, shown as points,
demonstrate a narrowing of resonance line-shapes as the sample moves
away from the FeCo tip indicating gradients exceeding $10^5$ T/m
within a 100 nm spacing \cite{Nazaretski:2008}.

\begin{figure}[tp]
\includegraphics[trim=8.1cm 13.5cm 7.3cm 13.1cm,clip, width=8.5cm, page=4]{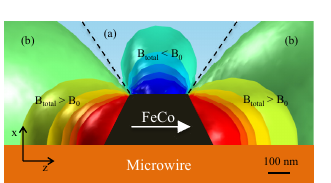}
\caption{ \label{fig4} Cross section through the center of the FeCo
tip with resonant slices in the xz-plane (surfaces of constant
$B_{total}$).  White arrow shows the magnetization direction of
FeCo. }
\end{figure}

In order to extract more detailed information from these resonances,
a simple theory is constructed from a magneto-static model of the
FeCo tip and an effective field model for adiabatic rapid passage in
the manner of the appendix in reference \cite{Degen:2009}.  The
model's input parameters include the geometry of the FeCo tip and
sample taken from scanning electron microscope (SEM) images, the
mechanical characteristics of the cantilever, and the form of the
adiabatic sweep waveforms \cite{SuppMat}. Using realistic
parameters, we see a good agreement between our model and the data,
especially considering our approximate knowledge of the morphology
of the sample.  The drop-like sample is modeled as a sphere with a
radius of 625 nm, close to the radius of curvature observed by SEM.
This agreement allows us to plot the likely field distribution
around the FeCo tip in Fig.~\ref{fig4}. In this figure, we show the
surfaces of constant field around the FeCo tip for a saturating
field ${\bf B}_0$ pointing along $\hat{z}$.  In MRFM, these regions
are also known as ``resonant slices'' since they describe the region
in space occupied by the spins resonant with a particular carrier
frequency $\omega_{RF}$.  $\omega_{RF}$ is determined by the
magnetic resonance condition $\omega_{RF} / \gamma = B_{total} =
|{\bf B}_0 + {\bf B}_{tip}|$, where $\gamma$ is the gyromagnetic
ratio of the isotope of interest, and ${\bf B}_{tip}$ is the field
provided by the FeCo magnet.  The form and position of a resonant
slice determines the point-spread function of an MRFM measurement
and thus is required for any imaging application \cite{Degen:2009}.

In Fig.~\ref{fig4}, we see that there are two types of resonant slice
regions: (a) where $B_{total} < B_0$, and (b) where $B_{total} > B_0$.
The experiments shown in Fig.~\ref{fig2} show that strong signal can
be observed both when the sample is in region (a) as in
Fig.~\ref{fig3}(a) and when the sample is in region (b) as in
Fig.~\ref{fig3}(b).  Note that the resonance signal in
Fig.~\ref{fig3}(a) appears at carrier frequencies lower than the
magnetic resonance condition at $B_0$ (represented by the dashed
line).  This result depends on the fact that during this experiment
the sample is always within region (a).  Thus $B_{total} < B_0$ for
all spins in the sample, resulting in magnetic resonance only for
frequencies less than the $B_0$ resonance.  Likewise, for a sample
always in region (b), such as shown in Fig.~\ref{fig3}(b), resonant
signal appears only for frequencies greater than the $B_0$ resonance.
Note that the small amount of resonant signal observed for frequencies
greater (less) than the $B_0$ resonance in Fig.~\ref{fig3}(a)
(Fig.~\ref{fig3}(b)) results from the small parts of the sample that
protrude into the adjacent region.  In both cases, as expected, the
amplitude of the spin signal increases as the sample moves closer to
the FeCo tip where magnetic field gradients are stronger.  Using the
magnetostatic model of Fig.~\ref{fig4} we can extrapolate that in both
regions (a) and (b), the maximum gradient relevant to MRFM ($\partial
B_{total} / \partial y$) exceeds $10^6$ T/m within 50 nm of the FeCo
tip.  More precise bounds on this gradient cannot be given because of
the approximate model of the morphology of the sample.  While
gradients in region (a) are always smaller than in region (b) for
similar spacings, both are comparable to gradients reported in the
most recent realizations of nanoscale MRFM.

\begin{figure}[bp]
\includegraphics[trim=7.9cm 14.0cm 7.6cm 13.5cm,clip, width=8.5cm, page=5]{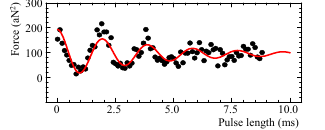}
\caption{ \label{fig5} Spin nutation at $T = 400$ mK. Force signal is
  measured as a function of pulse length. A $B_1$ of 12 mT
  is extracted from a decaying sinusoidal fit of the Rabi oscillations
  shown in red.}
\end{figure}

These experiments demonstrate that MRFM can be done both with a sample
positioned in the region above and in the region next to the FeCo tip.
Importantly, the positioning of the sample mounted on the end of the
cantilever next to the FeCo tip does not cause the cantilever to snap
in to contact even for spacings down to 50 nm.  The problem of snap
in, which is typically due to electrostatic and van der Waals forces,
is avoided here because while the cantilever is extremely soft along
$\hat{y}$, it is stiff along $\hat{z}$, and extremely stiff along
$\hat{x}$.

The accessibility of the different resonant slice regions depicted in
Fig.~\ref{fig4} allows for additional flexibility in future nano-MRI
imaging experiments.  The shape of the resonant slices and the
distribution of the magnetic field gradients determines the point
spread function for imaging and therefore the resolution of the
technique in each spatial direction.  In region (a), for instance,
strong gradients in $\hat{x}$ provide high spatial resolution along
this direction, while smaller gradients in the yz-plane reduce the
resolution in those directions.  Region (b), on the other hand, offers
large gradients and high resolution in $\hat{z}$ and reduced gradients
and resolution in the xy-plane.  The most recent high resolution
imaging experiments have positioned the sample directly above the FeCo
tip, achieving high spatial resolution in one direction and reduced
resolution in the remaining two.  By using the additional region next
to the FeCo tip, the high spatial resolution could also be achieved in
one more direction, allowing for a more complete high resolution
image.

In order to measure the magnitude of $B_1$ we apply the spin
nutation waveform described in reference \cite{PoggioAPL:2007,
SuppMat} to our microwire source.  Pulses of variable length are
inserted in the adiabatic sweep waveform every 500 cantilever cycles
(190 ms), resulting in the spin nutation signal shown in
Fig.~\ref{fig5}.  By fitting the data to a decaying sinusoid, we can
infer that the rotating RF magnetic field $B_1$ exceeds 12 mT (510
kHz for $^1H$) within 200 nm of the magnetic tip for $T < 500$ mK.
The correlation time $\tau_m$ of the spin signal was estimated from
the signal bandwidth and found to be on the order of 500 ms.  This
value is significantly longer than the $\sim 20$ ms observed for
$^1$H in previous experiments on organic material \cite{Degen:2009},
possibly due to the 3 times larger $B_1$ applied here
\cite{PoggioAPL:2007}. Such a long $\tau_m$ allows for the use of
pulse protocols that can improve the signal-to-noise ratio and
therefore the imaging resolution of nanoscale MRFM
\cite{Degen:2007}. The generation of large $B_1$ fields is also an
important technical achievement, since MRFM protocols utilizing
adiabatic rapid passage must satisfy the adiabatic limit $(\gamma
B_1)^2$ $\gg$ $\omega _{m} \Omega_{RF}$ in order to produce a
signal.  Here $\Omega_{RF}$ is the frequency modulation amplitude of
the adiabatic sweep waveform. Nuclear species with low $\gamma$
therefore require a large $B_1$ in order to be observable by MRFM.
In addition, nuclear resonances broadened by strain or other effects
require large $\Omega_{RF}$. Therefore this large measured $B_1$
amplitude indicates that nuclear MRFM on previously inaccessible
samples such as In or As nuclei in single self-assembled InAs
quantum dots, could be possible.

Assuming a magnetic field gradient of greater than $10^6$ T/m within 50
nm of our FeCo tip, and the low temperature force noise of our
cantilever near the surface of less than 10 aN/$\sqrt{\text{Hz}}$, we
estimate that our technique has a minimum detectable moment of $1.0
\times 10^{-23}$ J T$^{-1}$ Hz$^{-1/2}$ at this distance.  For In
moments in InAs self-assembled QDs, for example, this value represents
a sensitivity to the statistical polarization of an ensemble of $1.6
\times 10^4$ In moments in a typical integration time of one minute
\cite{Degen:2009}.  Given that there are between $10^4$ and $10^5$
In nuclei in a typical self-assembled QD, these parameters promise
that imaging of single QDs by MRFM should be possible.

\begin{acknowledgments}
 Data shown in Fig. 1 were taken in Dr. Dan Rugar's lab at IBM Almaden.
 We acknowledge support from the Canton Aargau, the Swiss NSF (grant 200021 1243894),
 and the Swiss Nanoscience Institute.
\end{acknowledgments}

%\pagebreak

\end{document}